\documentclass[aps,prl,reprint,showpacs,showkeys,nofootinbib,
superscriptaddress,preprintnumbers,groupedaddress]{revtex4-1}
\usepackage{amsmath,amssymb,bm,mathrsfs}
\usepackage{srcltx}
\usepackage{dsfont}
\usepackage{epsfig}
\usepackage{slashed}
\usepackage{bbold}
\usepackage{psfrag}
\usepackage{xcolor}
\PassOptionsToPackage{caption=false}{subfig}
\usepackage{subfig}
\usepackage{xfrac}
\usepackage{multirow}
\usepackage{booktabs}
\usepackage{hyperref}
\hypersetup{colorlinks=true,linkcolor=blue,citecolor=blue,urlcolor=violet}
\usepackage{relsize}

\newcommand{\be}{\begin{equation}}
\newcommand{\ee}{\end{equation}}

\newcommand{\beq}{\begin{equation}}
\newcommand{\eeq}{\end{equation}}
\newcommand{\ba}{\begin{array}}
\newcommand{\ea}{\end{array}}
\newcommand{\bea}{\begin{eqnarray}}
\newcommand{\eea}{\end{eqnarray} }
\newcommand{\bal}{\begin{align}}
\newcommand{\eal}{\end{align}}
\newcommand{\bi}{\begin{itemize}}
\newcommand{\ei}{\end{itemize}}
\newcommand{\ben}{\begin{enumerate}}
\newcommand{\een}{\end{enumerate}}
\newcommand{\bc}{\begin{center}}
\newcommand{\ec}{\end{center}}
\newcommand{\bt}{\begin{table}}
\newcommand{\et}{\end{table}}
\newcommand{\btb}{\begin{tabular}}
\newcommand{\etb}{\end{tabular}}

\newcommand{\eq}{{\rm{eq}}}

\def\({\left(}
\def\){\right)}

\usepackage{soul}

\begin{document}

\title{Heavy Thermal Relics from Zombie Collisions}

\author{Eric David Kramer}
\email{ericdavidkramer@gmail.com}
\affiliation{Racah Institute of Physics, Hebrew University of Jerusalem, Jerusalem 91904, Israel}
\author{Eric Kuflik}
\email{eric.kuflik@mail.huji.ac.il}
\affiliation{Racah Institute of Physics, Hebrew University of Jerusalem, Jerusalem 91904, Israel}
\author{Noam Levi}
\email{noam@mail.tau.ac.il}
\affiliation{Raymond and Beverly Sackler School of Physics and Astronomy, Tel-Aviv University, Tel-Aviv 69978, Israel}
\author{Nadav Joseph Outmezguine}
\email{Nadav.Out@gmail.com}
\affiliation{Raymond and Beverly Sackler School of Physics and Astronomy, Tel-Aviv University, Tel-Aviv 69978, Israel}
\author{Joshua T. Ruderman}
\email{ruderman@nyu.edu}
\affiliation{
Center for Cosmology and Particle Physics,
Department of Physics, New York University, New York, NY 10003, USA.
}

\begin{abstract}
We propose a new thermal freezeout mechanism which results in dark matter masses exceeding the unitarity bound by many orders of magnitude, without violating perturbative unitarity or modifying the standard cosmology. The process determining the relic abundance is  $\chi \zeta^\dagger \to \zeta \zeta$, where $\chi$ is the dark matter candidate. For $ m_\zeta < m_\chi < 3 m_\zeta$,  $\chi$ is cosmologically long-lived and scatters against the exponentially more abundant $\zeta$. Therefore, such a process allows for exponentially heavier dark matter for the same interaction strength as a particle undergoing ordinary $2 \to 2$ freezeout; or equivalently, exponentially weaker interactions for the same mass. We demonstrate this mechanism in a leptophilic dark matter model, which allows for dark matter masses up to $10^{9}$ GeV.
\end{abstract}

\maketitle

\noindent {\bf  Introduction:} 
The observational evidence for the existence and ubiquity of Dark Matter (DM) is well established, yet its origin and particle nature remains unknown. This puzzle has driven a several decades long exploration into the landscape of potential DM models and cosmological mechanisms for producing the relic abundance of DM\@.  Among these, the prospect of DM particles  thermally coupled to the Standard Model (SM) in the early Universe has been especially prominent. 

During the cosmological evolution of the Universe, at early times, the interaction rate between the SM and the dark sector is fast, keeping the two baths in chemical and thermal equilibrium. When the interaction rate falls below the Hubble expansion rate, chemical equilibrium between the SM and the dark sector ceases. Soon afterwards, the interactions completely stop and the abundance of DM is set by a process known as thermal freezeout.  The most widely explored paradigm representing this concept is the Weakly Interacting Massive Particle (WIMP), where the relic abundance is set by the freezeout of DM-DM annihilations to the SM\@. The WIMP is a particularly promising candidate, as it predicts DM masses at the weak scale with DM annihilation rates on the order of the weak interactions, thus relating DM production in the early Universe to theories of the weak scale and to new physics at the experimental frontier.

Within the WIMP scenario, there is an upper bound on the mass of DM set by perturbative unitarity of roughly $m_{\rm WIMP}\sim 100$~TeV~\cite{Griest:1989wd}. Are there minimal extensions to the WIMP paradigm that predict heavier DM masses than this unitarity bound, leading to qualitatively different experimental signatures? Attempts at answering this question have focused on out-of-equilibrium dynamics with the SM and/or non-standard cosmological histories~\cite{Hui:1998dc,Kolb:1998ki,Chung:1998rq,Chung:2001cb,Feng:2008mu,Harigaya:2014waa,Davoudiasl:2015vba,Randall:2015xza,Dev:2016xcp,Harigaya:2016vda,Berlin:2016vnh,Berlin:2016gtr,Bramante:2017obj,Berlin:2017ife,Hamdan:2017psw,Cirelli:2018iax,Babichev:2018mtd,Hashiba:2018tbu,Hooper:2019gtx,Davoudiasl:2019xeb}.  If DM is a composite object, such as a hadron, the unitarity bound applies to the size of the object and not the masses of its constituents~\cite{Griest:1989wd, Harigaya:2016nlg,Smirnov:2019ngs,Contino:2018crt,Gross:2018zha,Geller:2018biy}.  

Thermal freezeout mechanisms considering topologies beyond the WIMP have been considered~\cite{DEramo:2010keq,Hochberg:2014dra,Kuflik:2015isi,Cline:2017tka,DAgnolo:2017dbv,Smirnov:2020zwf}, but until recently none have evaded the unitarity bound. A thermal mechanism that exceeds the unitarity bound, without modifying the standard cosmology, was proposed for the first time in Ref.~\cite{Kim:2019udq}, requiring a chain of interactions but allowing DM masses as high as $10^{14}$~GeV\@. In this \textit{Letter} we present a new thermal 2-to-2 freezeout mechanism, that requires just two interactions, and allows for DM masses as heavy as $10^{10}\,\rm GeV$, without violating unitarity or modifying the standard cosmology. Conversely, this mechanism can achieve thermal weak scale DM, but with much smaller interaction rates than the WIMP\@.

Our setup consists of a DM candidate ($\chi$), a SM portal, and at least one extra interacting degree of freedom in the dark sector ($\zeta$).
The dark sector is in equilibrium with the SM at early times, maintained by $\zeta \zeta^\dagger$ annihilations to the SM, and internally in equilibrium via the process $\chi \zeta^\dagger \to \zeta \zeta$ between the two dark particles.  When the masses of the dark sector particles obey the hierarchy $m_\zeta < m_\chi$, heavy DM naturally arises in our setup, as we detail below. The reason is that the mass splitting causes $\zeta$ to be exponentially more abundant than $\chi$ in chemical equilibrium, allowing for $\chi$ to be removed by scattering efficiently against a particle that is exponentially more abundant than itself. Note that a similar process was considered in Ref.~\cite{Berlin:2017ife} for heavy DM within a non-thermal and non-standard cosmological history; in contrast, here we show how to obtain heavy DM thermally and within a standard cosmology. \\

\noindent {\bf  General Idea:} 
Consider a DM particle, $\chi$, whose number density changes at early times via an interaction of the form
\begin{equation}
\label{eq:int}
{\cal L}\supset \chi^\dagger  \zeta \zeta \zeta,
\end{equation}
 with some field $\zeta$.  The DM number density can deplete via the process 
\begin{equation}
	\chi \zeta^\dagger  \to \zeta \zeta\, . \label{exppro}
\end{equation}
This process behaves similar to a zombie infection~\cite{PhysRevE.92.052801}, where a $\zeta$ particle (the zombie), infects the DM $\chi$ (a survivor) and turns it into a zombie.  The final DM abundance consists of the survivors, $\chi$, that persist after the process of Eq.~\eqref{exppro} decouples (the outbreak ends; fortunately).
We refer to this process as a  \emph{zombie collision}.   

 We restrict ourselves to $m_\zeta < m_\chi < 3 m_\zeta$, where such interactions do not induce (on-shell) $\chi\to \zeta \zeta 	\zeta$ decays, and the process in Eq.~\eqref{exppro} does not enter a  forbidden regime~\cite{Griest:1990kh,DAgnolo:2015ujb}.  
Whenever $m_\chi > m_\zeta$, the processes in Eq.~\eqref{exppro} can maintain chemical equilibrium longer than if the $\chi$ particle was annihilating with another $\chi$, 
because the interaction rate is proportional to $n_\zeta$, which is exponentially larger than $n_\chi$ when $\chi$ becomes non-relativistic.  
Thus, if this process is responsible for $\chi$ freezeout, the correct relic abundance is obtained for smaller interactions than WIMP-like DM, for the same DM mass. Similarly, since zombie collisions can be very efficient, this scenario also allows for heavier DM than the WIMP unitarity bound, without violating unitarity.

\begin{figure}[t!]
\begin{center}
\includegraphics[width=0.48 \textwidth]{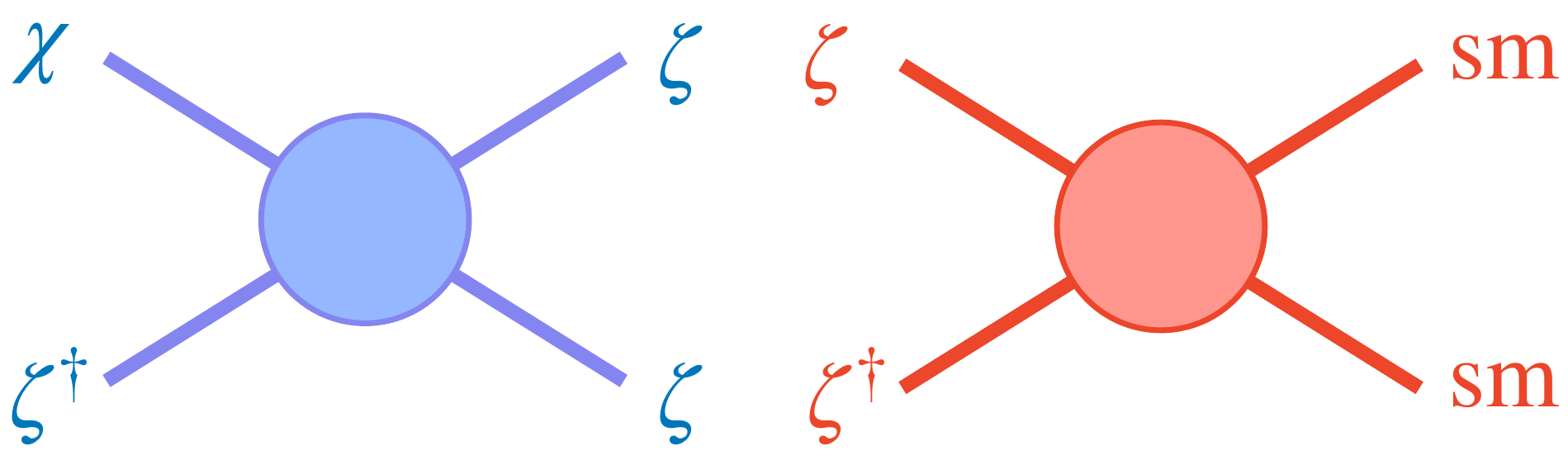}
\end{center}
\caption{\small 
Schematic representation of the processes setting the relic abundance. On the left, $\chi$ is our DM candidate and $\zeta$ is a hidden sector zombie particle that turns a $\chi$ into a $\zeta$. The process on the right maintains chemical equilibrium between the dark sector and the SM\@.
\label{fig:cartoon}}
\end{figure}

 \begin{figure*}[t!]
\begin{center}
\includegraphics[width=0.48 \textwidth]{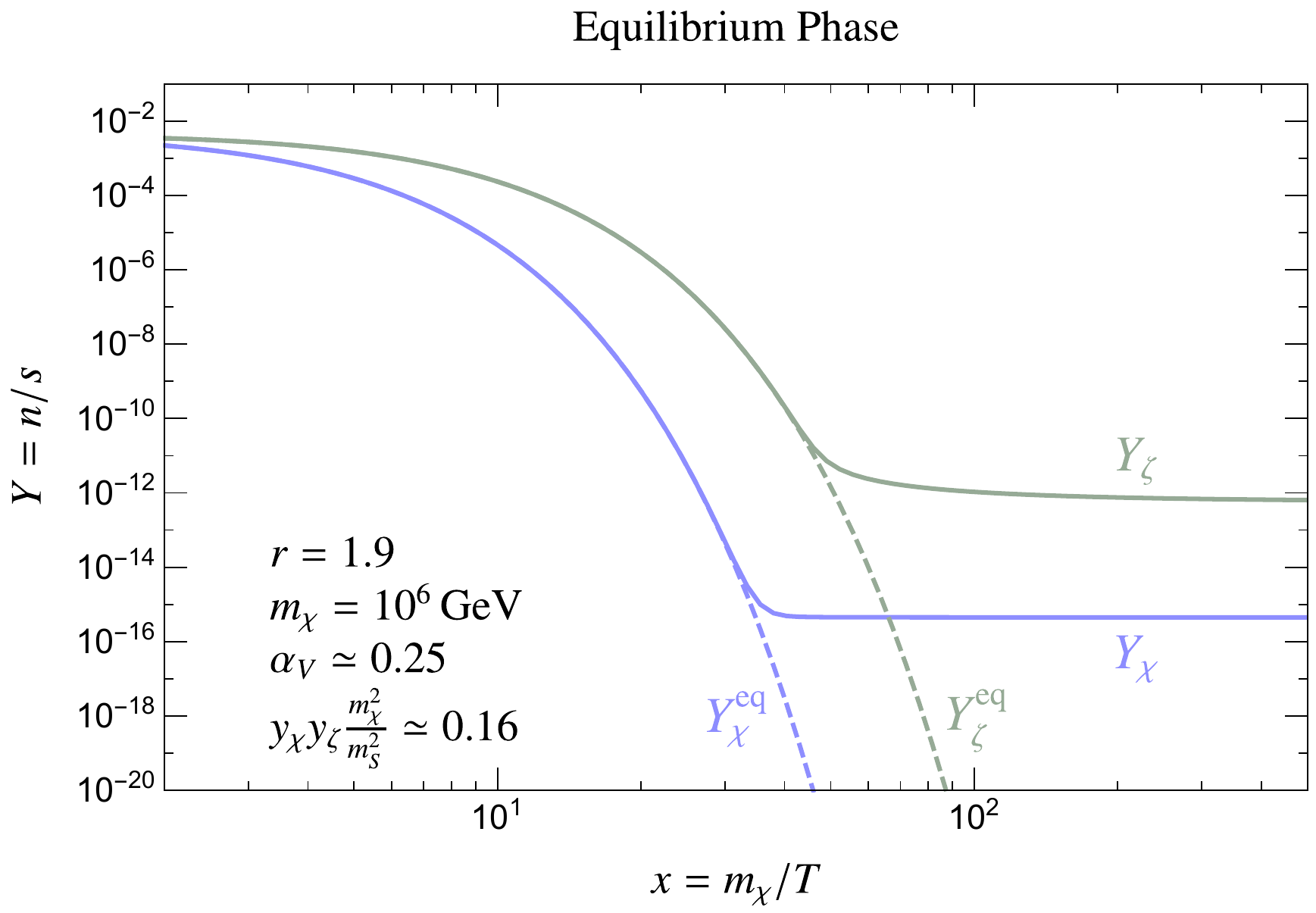}\hspace{0.04\textwidth}\includegraphics[width=0.48  \textwidth]{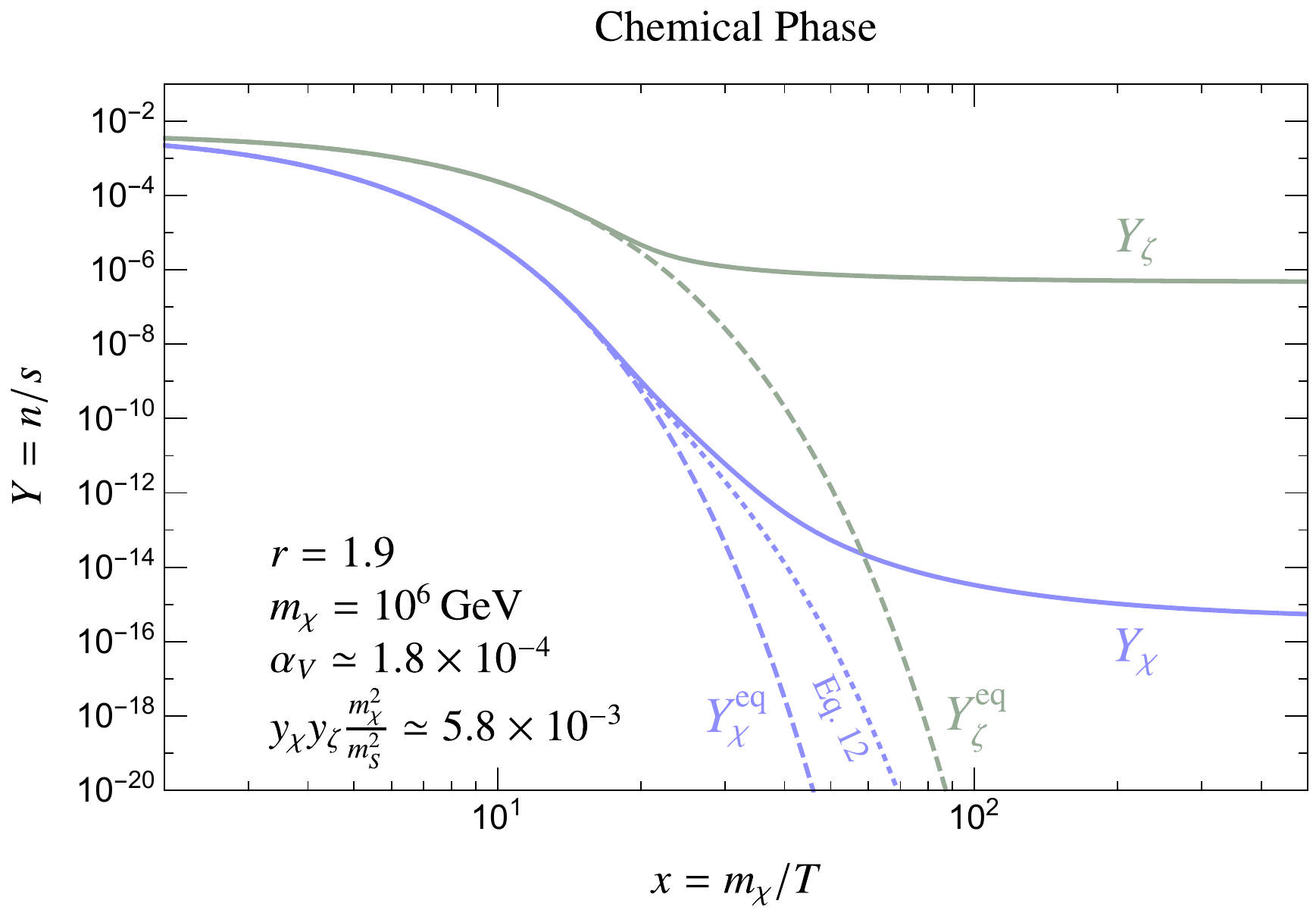}
\end{center}
\caption{\small 
Thermal evolution of the number density of $\chi$ {\bf(blue)},  $\zeta$ {\bf(green)}, and the zero chemical potential equilibrium distribution (\textbf{dashed lines}). \textit{Left - Equilibrium Phase:} $\zeta$ is in chemical equilibrium with the SM bath when $\chi$ freezes out. \textit{Right - Chemical Phase:} $\zeta$ has departed chemical equilibrium with the SM right before $\chi$ freezes out. Both panels show the evolution of yield for dark sector particles, with $m_\chi=1.9m_\zeta= 10^6\,\rm GeV$ for interaction rates producing the observed DM relic abundance. The right panel  also shows \textbf{(dotted line)} the $\chi$  equilibrium distribution after it develops a chemical potential, as described above Eq.~\eqref{eq:new_n}. \label{fig:relic}}
\end{figure*}

To realize this mechanism we consider the two processes shown in Fig.~\ref{fig:cartoon}, 
\bea
\chi \zeta^\dagger	\to \zeta \zeta ,
&    ~~~~~ ~~~~~ &
  \zeta \zeta^\dagger \to \rm sm  \, \rm sm \, ,
\eea
where `sm' is a light particle which is either part of the SM bath or thermalized with it. The possibility of other interactions, such as $\chi\chi^\dagger$ annihilations, will be discussed below in the context of a specific UV model.  
Coscattering~\cite{DAgnolo:2017dbv,Kim:2019udq,Garny:2017rxs,DAgnolo:2019zkf} is a different example where the DM abundance can be set by the decoupling of 2-to-2 scattering of DM against a lighter state.

The Boltzmann equations governing the evolution of the $\chi$ and $\zeta$ number densities are given by 
\beq
	\dot{n}_{\chi}+3 H n_{\chi} =-	 \langle\sigma_{\chi \zeta \to \zeta\zeta} v\rangle\left(n_{\chi} n_{\zeta} -	n_{\zeta}^2\frac{ n_{\chi}^{\rm eq} }{n^{\rm eq}_{\zeta}}\right)\,, \label{eq:BEchi}\eeq
	\vspace{-4.5mm}
\beq	\dot{n}_{\zeta}+\dot{n}_{\chi} +3 H (n_{\zeta}+n_{\chi}) = -\langle\sigma_{\zeta \zeta \to \rm sm \, sm } v\rangle \left(n_{\zeta}^2 -	{n_{\zeta}^{\rm eq}}^2  \right)\!,~~ \hspace{-3mm}
	\eeq
where  the superscript `$\rm eq$' denotes equilibrium abundances at zero chemical potential. 
The density of $\chi$ will depart from its equilibrium distribution (and freeze out soon after) when the rate of  $\chi$ number-changing process drops below the Hubble expansion rate. This happens approximately when~(see e.g. Eq.~(5.40) in~\cite{Kolb:1990vq})
\begin{equation}\label{eq:chi_fo}
n_\zeta(x_{ \chi}) \langle\sigma_{\chi \zeta \to \zeta\zeta} v\rangle = x_{ \chi}\,H(x_{\chi})\,,
\end{equation} 
where we have defined $x \equiv m_\chi/T$ and $x_{ \chi}$ is the temperature when $\chi$ departs equilibrium. Eq.~\eqref{eq:chi_fo} determines this temperature,  which will be used to estimate the $\chi$ relic abundance.   
Unlike the WIMP, the freezeout dynamics depends on whether the zombies, $\zeta$, follow an equilibrium distribution or instead have already frozen out from the thermal bath by the time $\chi$ departs equilibrium with the $\zeta$. 
In what follows we describe these possibilities for the different phases of dark sector freezeout. \\

\noindent{\it Equilibrium Phase -- Zombies in equilibrium throughout freezeout:} 
When the $\zeta \zeta^\dagger \leftrightarrow \rm sm  \, \rm sm$ interactions are efficient in maintaining equilibrium of $\zeta$ with the SM bath, $\chi$ evolves 
according to the Boltzmann equation 
\begin{equation}
	\dot{n}_{\chi}+3 H n_{\chi} =-	 n^{\rm eq}_{\zeta}\langle\sigma_{\chi \zeta \to \zeta\zeta} v\rangle\left(n_{\chi} -	n_{\chi}^\eq\right) \, .
\end{equation}
The relic abundance can be estimated using the instantaneous freezeout approximation, utilizing the fact that at freezeout the $\chi$-$\zeta$ system is still in equilibrium:
\begin{equation}
n_\chi(x) \simeq r^{3/2}  \exp\left[-\frac{r-1}{r}x\right]n_\zeta(x) ,
\end{equation}
where we defined $r\equiv m_\chi/m_\zeta>1$. The rest of the relic abundance calculation proceeds via standard techniques. Parameterizing the cross-section as $\langle\sigma_{\chi \zeta \to \zeta\zeta} v\rangle    \equiv \alpha_\chi^2/m_\chi^2$, the DM mass required to match the observed abundance is
\beq
m_\chi \simeq \left((\alpha_\chi^2 m_{\rm pl})^r T_{\rm eq}\right)^{\frac{1}{1+r}}\,,\label{eq:phase1}
\eeq
where $T_{\rm eq}\simeq 0.8$~eV is the temperature at matter radiation equality and $m_{\rm pl} $ is the 
Planck mass. The same expression with $r=1$ is the known relationship from the standard WIMP calculation. The gain in DM mass over the WIMP is evident from the equation above---the relic calculation puts an exponentially larger weight on $m_{\rm pl}$ vs.~$T_{\rm eq}$, leading to exponentially heavier DM for the same interaction strength (or conversely, exponentially smaller interaction strength for the same size DM mass). The thermal evolution of this phase, for the specific model we describe later, is shown in the left panel of Fig.~\ref{fig:relic}.

\begin{figure}[t!]
	\begin{center}
	\includegraphics[width=0.48 \textwidth]{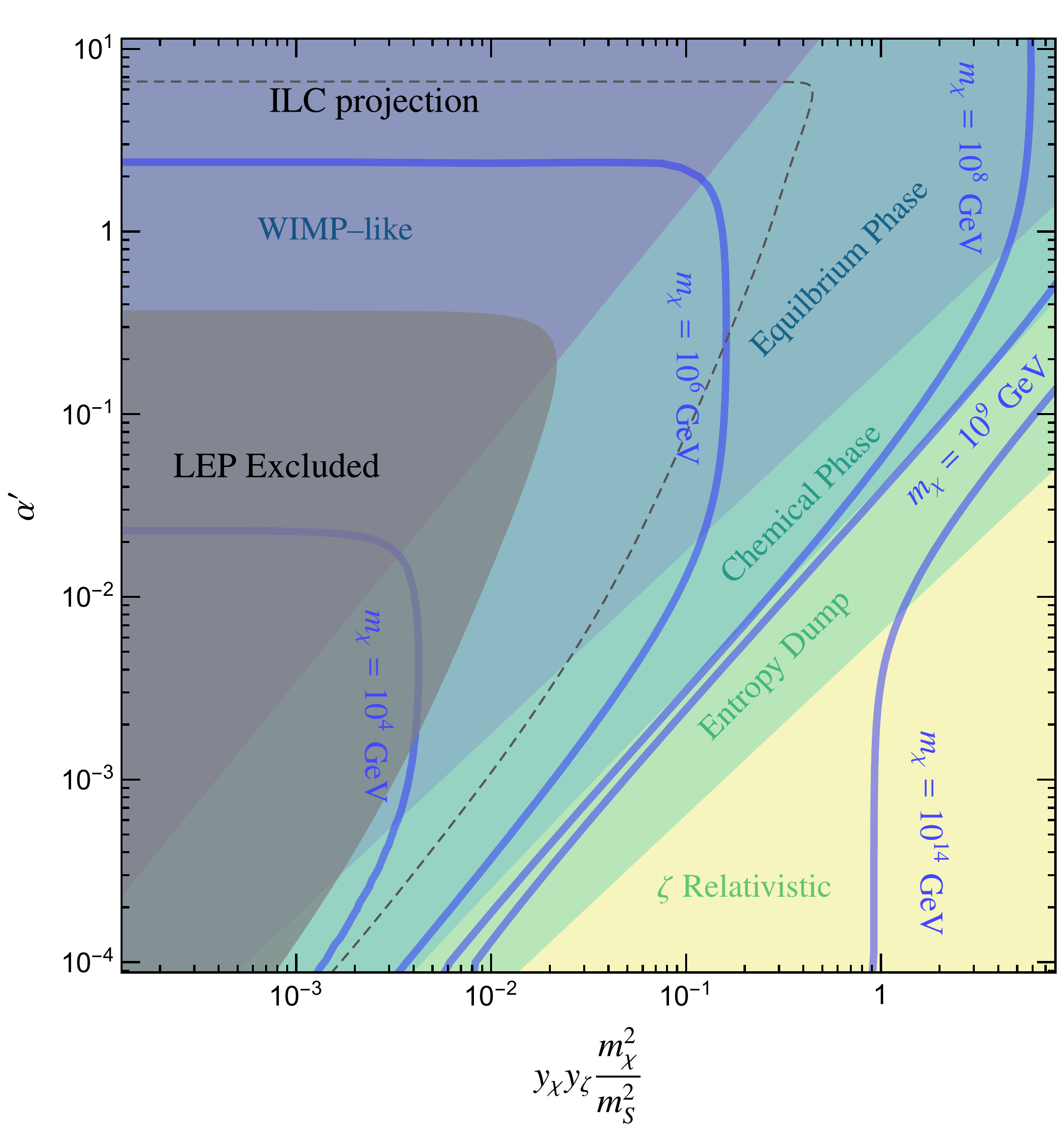}
	\caption{\small $\alpha^\prime$ vs $y_\chi y_\zeta m_\chi^2 /m_S^2$, with contours of constant mass required to match the observed relic abundance. We fix the ratios $m_\chi=1.9\,m_\zeta$ and $m_{Z^\prime}= 0.1 m_{\chi}$. We take the lifetime of $\chi$ to be $\tau_\chi=10^{27}$~sec. Differently colored  regions correspond to different phases of the model. See the main text for an explanation of each phase. The \textbf{shaded gray} region is excluded by LEP $Z'$ searches, while \textbf{dashed gray} indicates projected ILC sensitivity for $Z'$ searches. \label{fig:phase}}
	\end{center}
\end{figure}

In most models, one would expect $\chi \chi$ to also annihilate to the SM or zombies $\zeta$. If these annihilations are faster at freezeout than $\chi \zeta^\dagger \to \zeta\zeta$, then the DM will behave as a standard WIMP\@. Parameterizing the annihilation cross section as $ \left<\sigma_{\chi \chi \to \rm sm\,sm} v\right> = \alpha_{\rm wimp}^2/m_\chi^2$, the condition for zombie collisions to control the abundance is easily determined by comparing Eq.~\eqref{eq:phase1} to the analogous equation for the WIMP:
\beq
\left((\alpha_\chi^2 m_{\rm pl})^r T_{\rm eq}\right)^{\frac{1}{1+r}} \gtrsim \alpha_{\rm wimp}\left(m_{\rm pl} T_{\rm eq}\right)^{\frac{1}{2}} \,. \label{eq:boundary1}
\eeq
This equation gives the approximate phase boundary between the WIMP and the Equilibrium Phase, which is shown in Fig.~\ref{fig:phase} for the model realization we discuss below.\\

\noindent{\it Chemical Phase -- Zombies develop chemical potential:} 
It is possible that $\zeta$ freezes out of equilibrium with the SM bath, before $\chi$ decouples from $\zeta$.  In this case, $\zeta$  has a constant comoving abundance when $\chi$ departs from equilibrium with it. (A non-zero chemical potential of another state also impacts DM freezeout in Refs.~\cite{Bandyopadhyay:2011qm,Farina:2016llk,Dror:2016rxc,Cline:2017tka,Berlin:2017ife}.) This phase requires a more subtle treatment, since the sudden freezeout approximation will not give a good estimate of the relic abundance.  We show here a qualitative analysis of this phase; a more detailed quantitative analysis appears in the Appendix.

The $\zeta$ distribution departs from equilibrium with the SM bath  at a temperature $x_\zeta$,  defined by 
\begin{equation}
	n_\zeta^{\rm eq}(x_\zeta)\langle\sigma_{\zeta \zeta \to \rm sm \, sm } v\rangle=x_\zeta H(x_\zeta)\, . \label{eq:psi_fo}
\end{equation}
At $x>x_\zeta$ the comoving number density of $\zeta$ begins to freeze  out and soon approaches a constant value. The evolution of the $\chi$ density is then calculated by plugging the frozen-out  density of $\zeta$ into Eq.~\eqref{eq:BEchi}.
After $\zeta$ freezeout, the $\chi$ distribution  begins to trace a new equilibrium distribution with a chemical potential dictated by the frozen out abundance of $\zeta$, 
\begin{equation}\label{eq:new_n}
n_\chi^{\rm eq} \to  n_\zeta\frac{n_\chi^{\rm eq}}{n_\zeta^{\rm eq}} \,.
\end{equation}
Here $n_\zeta$ is the frozen out abundance of $\zeta$ that evolves approximately as $n_\zeta \sim n_\zeta(x_\zeta) (x_\zeta/x)^3$. 
Following similar steps to those leading to Eq.~\eqref{eq:phase1}, we show in the Appendix~\ref{sec:appendix}that
\begin{equation}
	m_\chi\sim\left[(\alpha_\zeta^2 m_{\rm pl})^{r+\Delta} T_{\rm eq}\right]^{\frac{1}{1+r+\Delta}}\,,\label{eq:phase2}
\end{equation}
where we have defined $\langle\sigma_{\zeta \zeta \to \rm sm \, sm } v\rangle    \equiv \alpha_\zeta^2/m_\chi^2$ and
\begin{equation}\label{eq:Delta}
	\Delta=r\frac{x_\chi}{x_\zeta}+(r-1)\left(\frac{x_\chi}{x_\zeta}-1\right)>0 \, ,
\end{equation}
with $x_\chi$ and $x_\zeta$ as defined in Eq.~\eqref{eq:chi_fo} and Eq.~\eqref{eq:psi_fo}, respectively. 
We see that the Chemical Phase leads to even higher DM masses than the Equilibrium Phase, as can be seen by comparing Eq.~\eqref{eq:phase2} to Eq.~\eqref{eq:phase1} . The thermal evolution for this phase is plotted in the right panel of Fig.~\ref{fig:relic}, for the model realization we present below.

The crossover between the Equilibrium Phase and the Chemical Phase occurs when the rate for $\chi$ to undergo $\chi\zeta^\dagger \to \zeta \zeta $ is approximately the same as the rate for $\zeta$ to annihilate via $\zeta\zeta^\dagger \to \rm sm ~ sm$ at freezeout. This is simply the condition:
\beq
\alpha_\zeta \simeq \alpha_\chi\,.
\eeq

\noindent{\it Additional phases:} 
Since the zombies can freeze out with a large abundance, they can come to dominate the energy of the Universe, leading to an early period of matter domination. If this happens, the $\zeta$ must decay and reheat the radiation bath. The large entropy dump from the decay will change the relic density calculation. We discuss this more in the following section. Finally, if $\zeta$ freezes out when relativistic, the model enters a new phase where the relic abundance of $\chi$ no longer depends on when $\zeta$ freezes out. Additionally, the $\chi$ temperature may not match the SM bath temperature, but the temperatures can be comparable. These two possibilities---non-thermal DM and large dilution via entropy dump---were studied in a similar framework in Ref.~\cite{Berlin:2017ife}. A main goal of this \textit{Letter} is to demonstrate the possibility of very heavy thermal DM within perturbative unitarity, without modifying early cosmology. 	For this reason, we leave detailed discussion of  these additional phases for future work.
\\

\noindent {\bf  Unitarity and DM decays:} 
The zombies, $\zeta$, are either stable, and themselves constitute a component of dark matter in addition to $\chi$, or are unstable.  If the zombies are stable, they must satisfy the unitarity bound, applied to $\zeta \zeta \to \rm sm\, sm$, and therefore $\chi$ is at most $\mathcal{O}(1)$ heavier by the assumption that $m_\chi < 3 m_\zeta$.  If the zombies are unstable, then both $\zeta$ and $\chi$ can have masses that far exceed the unitarity bound.  In this case the abundance of $\zeta$ will exceed the abundance of $\chi$ at freezeout, but this energy density is removed by decays of $\zeta$.  For the remainder of this letter we focus on unstable zombies.  

There are two important phenomenological consequences of the fact that $\zeta$ freezes-out with an abundance larger than $\chi$ and subsequently decays. 
 First, if $\zeta$ decays, then $\chi$ is unstable via the process $\chi \to \zeta \zeta \zeta$, where some or all of the $\zeta$'s are produced off-shell and decay. This leads to potentially strong indirect detection signatures and constraints.   Indirect detection constraints on the DM decay lifetime $\tau_\chi$, e.g.~from the diffuse gamma-ray spectrum, can be as strong as $\tau_\chi\gtrsim  10^{27}~\rm sec$~\cite{Cirelli:2012ut,Essig:2013goa,Blanco:2018esa}.

Second, if $\zeta$ freezes out with a large abundance, it can come to dominate the energy of the Universe before it decays away. The large entropy dump that accompanies the decay effectively dilutes $\chi$ (see Eq. (5.73) of Ref.~\cite{Kolb:1990vq}), allowing for DM masses beyond the WIMP unitarity bound.  Additionally, a sufficiently long-lived and abundant $\zeta$ can imply that the Universe was matter dominated during BBN, spoiling the successful predictions of the standard cosmological scenario.\\

\noindent {\bf Example Model:} 
Consider a model, where the SM is extended with a gauged U(1)$_{e- \mu}$ lepton number and a dark sector containing fermions, $\chi$  and $\zeta$, and a scalar field~$S$, with U(1)$_{e- \mu}$ charges $q_\zeta= 1$, $q_\chi= 3$, and $q_S= -2$.  The most general renormalizable (and parity invariant) Yukawa interactions are
\begin{equation}
		\mathcal{L}_{\rm int} = y_{\zeta }S \bar\zeta^c \zeta + y_{\chi }S \bar\zeta \chi + y_{e}  H \bar\zeta L_e  + y_{\mu}  H \bar\zeta^c L_\mu +\mathrm{h.c.}
\end{equation}
where $H$ is the Higgs doublet and $L$ is the lepton doublet. We take tree-level  kinetic mixing between the new U(1) and hypercharge to be absent, but kinetic mixing is radiatively generated by $e$ and $\mu$ loops. The Yukawa couplings generate zombie collisions, $\chi \zeta^\dagger \to \zeta \zeta$, while the gauge interactions generate $\zeta \zeta^\dagger\to \rm sm \, \rm sm$, where `sm' here can be a SM lepton or the $e- \mu $ gauge boson. The Yukawa and gauge interactions also generate $\chi \bar \chi$ annihilations, which if responsible for $\chi$ freezeout, will lead to a WIMP-like scenario.
For other models of leptophilic DM, see for example Refs.~\cite{Fox:2008kb,Chao:2010mp,Chang:2014tea,Bai:2014osa,Bell:2014tta,Schwaller:2013hqa,Kile:2014jea,Freitas:2014jla,Fornal:2017owa,Rawat:2017fak,Madge:2018gfl,Duan:2017qwj,Banerjee:2018mnw}.

In Fig.~\ref{fig:phase}  we show curves of constant DM relic abundance  $\Omega_\chi \simeq 0.27$~\cite{Aghanim:2018eyx} and the different phases of freezeout.  Here we take $m_\chi=1.9\,m_\zeta$ and $m_{Z^\prime}= 0.1 m_{\chi}$, where $Z^\prime$ is the massive U(1)$_{e- \mu}$ gauge boson. Note that for smaller values of the vector mass $m_{Z^\prime}$, the 
various cross sections will be Sommerfeld enhanced due to the U(1)$_{e -\mu}$ force, leading to even heavier DM for the same coupling strength~\cite{ArkaniHamed:2008qn}.  We also take $m_S \gtrsim \mathcal{O}({\rm few})(m_\chi+m_\zeta)$  so that  $\chi \bar\zeta \to \zeta \zeta $ is not on the $S$-resonance and can be approximated by a contact interaction.   We fix the lifetime of $\chi \to \zeta H H L L $ to be $\tau_\chi =10^{27}$~sec, which is calculated using {\tt FeynRules}~\cite{Alloul:2013bka} and {\tt Madgraph}~\cite{Alwall:2014hca}. 

The results of the thermal evolution of our example model contains different phases, depending on the relative size of the interactions. For large U(1)$_{e- \mu}$ gauge coupling, $\alpha^\prime$, the annihilations of $\chi \bar\chi$ pairs is very efficient, and the DM behavior is WIMP-like. This is seen in the top left region of Fig.~\ref{fig:phase}, labeled `\texttt{WIMP-like}', where constant mass lines are horizontal since the relic abundance is insensitive to the Yukawa interactions. As $\alpha^\prime$ drops, the $\chi \zeta^\dagger \to \zeta \zeta $ process becomes more efficient  at late times. If the gauge coupling is still large enough to maintain chemical equilibrium between the dark and visible sectors until $\chi$ freezeout, then DM enters the Equilibrium Phase, corresponding to the vertical lines in the region labeled `\texttt{Equilibrium Phase}' in Fig.~\ref{fig:phase}. 
If $\zeta$ freezes out before $\chi$ freezes out, the DM will be in the Chemical Phase. This phase is further separated into 3 regions. The first, labeled `\texttt{Chemical Phase}', signals  when $\zeta $ freezes out when non-relativistic, but never dominates the energy density of the Universe. The second region, labeled `\texttt{Entropy Dump}', shows where $\zeta$ freezes out non-relativistically, but dominates the energy density before it decays, leading to a non-negligible dilution of the DM density. The final region, labeled `\texttt{$\boldsymbol \zeta$ Relativistic},' shows where $\zeta$ freezes out while still relativistic. \\

\noindent {\bf Signatures and Constraints:}  
We now discuss the generic phenomenological signals of the mechanism and signatures and constraints of the leptophilic model presented above. Since the focus of this paper is on heavy DM, below we only discuss parameter regions that correspond to $m_\chi\gtrsim 100~\rm  GeV$. We  leave lighter DM phenomenology within this framework for a future study.
 
 The mechanism generically predicts an indirect detection signal from dark sector decays. The $\zeta$ particles must decay before Big Bang Nucleosynthesis (BBN) so as not to obstruct light element formation~\cite{Ellis:1984er,Kawasaki:2017bqm}. On the other hand, a decay of $\zeta$  induces a decay of $\chi$ through the interaction Eq.~\eqref{eq:int}. Late time $\chi$ decays may produce ultra high energy cosmic rays (UHECR), detectable by diffuse gamma ray satellites~\cite{Ackermann:2014usa}, high energy neutrino experiments~\cite{Abbasi:2011ji}, and in dedicated UHECR observatories~\cite{Abreu:2011zze}. Combined with the BBN bound, the decay rate of $\zeta$ must reside in the window
\begin{align}
	 H_{\rm BBN} \lesssim \Gamma_{\zeta} \lesssim \Gamma_{\rm max} \, ,
\end{align}
where $\Gamma_{\rm max}$ is the value of $\Gamma_\zeta$ such that the lifetime of $\chi$ is within indirect detection bounds. 
In Fig.~\ref{fig:phase} we fix  $\Gamma_\chi^{-1}= (10^{27}~{\rm sec})^{-1}$. For the plotted parameter ranges,  $\Gamma_{\rm max}> H_{\rm BBN}$ everywhere, showing the possibility of a large indirect detection signal for all masses. 

The portal between the sectors leads to collider signatures. In Fig.~\ref{fig:phase}, we show the constraints from the measurement of the differential cross-section of lepton pairs at LEP~\cite{Schael:2013ita}, which bounds $m_{Z^\prime}/\sqrt{\alpha^\prime}  > 25$~TeV, and as well as future projections for the ILC with reach $m_{Z^\prime} /\sqrt{\alpha^\prime} > 200$~TeV~\cite{Freitas:2014jla}.  Dark production at the LHC for our model is suppressed because there is no tree-level interaction with quarks. However, due to the long-lived nature of $\zeta$, the dark sector may be discoverable in experiments designed to look for long-lived particles at the LHC, such as AL3X~\cite{Gligorov:2018vkc},  CODEX-b~\cite{Gligorov:2017nwh}, FASER~\cite{Feng:2017uoz}, and MATHUSLA~\cite{Curtin:2018mvb}. Finally, if a gauged U(1)$_{\rm B-L}$ is considered instead of the leptophilic model, the freezeout would remain unchanged, but there would be stronger collider signatures. We leave the study of a such a model to future work.  

Thus far, we have set the  tree-level gauge kinetic mixing between $Z'$ and hypercharge to zero. DM-proton scattering is generated at one-loop  by the electron and muon with cross section given by~\cite{Holdom:1985ag,Arcadi:2017hfi,Duan:2017qwj,Blanco:2019hah}
\begin{equation}
\sigma_p=\frac{64\mu_{\chi p}^2}{\pi m_{Z^\prime}^4}\alpha_{\rm em}^2\alpha^{\prime 2}\log^2\left(\frac{m_e}{m_\mu}\right) \, ,
\end{equation}
where $\mu_{\chi p} $ is the reduced DM-proton mass. 
Although this strength of interaction might seem relevant for direct detection experiments, we find that current nucleon recoil direct detection constraints from XENON1T~\cite{Aprile:2017iyp} do not outperform LEP $Z'$ searches. In the lower $\chi$ mass end of our model ($m_\chi\lesssim \rm GeV$), electron recoil experiments might be relevant and one would have to take into account the effect of shielding by the Earth~\cite{Emken:2019tni}, which in our case is dominated by a loop induced interaction with protons.\\

\noindent {\bf Conclusions:}  
 In this {\it Letter}, we presented a new 2-to-2 freezeout mechanism for thermally produced DM, allowing for very heavy DM without altering the standard cosmology. The DM relic abundance is set via zombie collisions in the dark sector: $\chi \zeta^\dagger \to \zeta \zeta$. We have shown that the exponential depletion of DM induced by these interactions can naturally lead to DM masses as high as $m_\chi\sim10^{10}~\rm{GeV}$, without violating perturbative unitary. This is achieved with only two  interactions, but a chain of zombie collisions, similar  to Ref.~\cite{Kim:2019udq},  can potentially allow for even heavier DM masses.  Additionally, higher DM masses may be realized within this framework when considering the dark sector to be asymmetric, relating leptogenesis to DM freezeout. We leave exploration of these possibilities for future work.\\

\begin{acknowledgements}
\noindent {\bf \em Acknowledgements.---} We would like to thank Asher Berlin, Timothy Cohen, Timon Emken, Rouven Essig, Michael Geller,  Roni Harnik, Yonit Hochberg, Hyungjin Kim, Gordan Krnjaic, Sam McDermott, Mukul Sholapurkar  and Tomer Volansky for useful discussions.  The work of EDK is supported in part by the Zuckerman STEM Leadership Program.  The work of EK is supported by the Israel Science Foundation (grant No.1111/17), by the Binational Science Foundation (grant No. 2016153) and by the I-CORE Program of the Planning Budgeting Committee (grant No. 1937/12). NJO is grateful to the Azrieli Foundation for the award of an Azrieli Fellowship. NL would like to thank the Milner Foundation for the award of a Milner Fellowship. JTR is supported by NSF CAREER grant PHY-1554858 and NSF grant PHY-1915409.  EDK, EK and NJO would like to acknowledge the GGI Institute for Theoretical Physics for enabling us to complete a significant portion of this work.   JTR acknowledges hospitality from the Aspen Center for Physics, which is supported by the NSF grant PHY-1607611.
\end{acknowledgements}

\appendix

\section{Appendix}
\label{sec:appendix}
In this appendix we derive an analytic understanding of the Decoupled Phase, when $\zeta$ departs chemical equilibrium before $\chi$ freezes out. The analysis here is made much simpler by defining the yield $Y=n/s$ , the number density normalized by the entropy density. We use  $x=m_\chi/T$ as a clock.   In this language the Boltzmann equations for $\chi$ and $\zeta$ take the form~\cite{Kolb:1990vq}
\begin{align}
	\frac{d Y_{\chi}}{d x}&=-\frac{\lambda_\chi}{x^{2}} Y_{\zeta}\left(Y_{\chi}-Y_{\chi}^{\mathrm{eq}} \frac{Y_{\zeta}}{Y_{\zeta}^{\mathrm{eq}}}\right),\\
	\frac{d Y_{\zeta}}{d x}+\frac{d Y_{\chi}}{d x}&=-\frac{\lambda_\zeta}{x^{2}}\left(Y_{\zeta}^2-\left(Y_{\zeta}^{\mathrm{eq}}\right)^2\right),
\end{align}
with 
\begin{equation}
	\lambda_\chi =\frac{ \langle\sigma_{\chi \zeta \to \zeta\zeta} v\rangle s(m_{\chi})}{H(m_{\chi})}, ~~	\lambda_\zeta =\frac{ \langle\sigma_{\zeta \zeta \to \rm sm \, sm } v\rangle s(m_{\chi})}{H(m_{\chi})}
\end{equation}
and the thermally averaged cross-sections are defined as $\langle\sigma_{\chi \zeta \to \zeta\zeta} v\rangle    \equiv \alpha_\chi^2/m_\chi^2$ and $\langle\sigma_{\zeta \zeta \to \rm sm \, sm } v\rangle    \equiv \alpha_\zeta^2/m_\chi^2$. Above we assumed that the number of relativistic  degrees of freedom is constant and defined $s(m_\chi)$ and $H(m_\chi)$ as the entropy density and the Hubble parameter, respectively, evaluated at $T=m_\chi$. 
We are interested in cases where $\chi$ freezes out after $\zeta$ departs from chemical equilibrium with the SM bath. This, by virtue of the Boltzmann equations above, happens if $\alpha_\chi\gtrsim \alpha_\zeta$. In that case the freeze out of $\zeta$ is well described by a sudden freeze out, which happens at $x_\zeta$ defined through
\begin{equation}
	\frac{\lambda_\zeta Y_\zeta^{\rm eq}(x_\zeta)}{x_\zeta^2}\simeq 1 \, .
\end{equation}
For $x>x_\zeta$, the inverse process  ${\rm sm \, sm}\to\zeta\zeta $ stops,  and $Y_\zeta$  evolves according to 
\begin{equation}
	\frac{d Y_{\zeta}}{d x}\simeq-\frac{\lambda_\zeta}{x^{2}}Y_{\zeta}^2 \, .
\end{equation}
Integrating this equation one finds
\begin{equation}\label{eq:new_Y}
	Y_\zeta(x>x_\zeta)\simeq	\frac{x}{-x_{\zeta} +(x_\zeta +1)x} Y_{\zeta}(x_\zeta) \,  .
\end{equation}
After $\zeta$ departs form its equilibrium distribution, $\chi$ follows a new equilibrium distribution given by (similarly to  Eq.~\eqref{eq:new_n})
\begin{equation}
	\tilde{Y}_\chi(x)=Y_\zeta(x>x_\zeta)\frac{Y_\chi^{\rm eq}(x)}{Y_\zeta^{\rm eq}(x)} \, .
\end{equation}
This can be understood as a thermal distribution, but with a chemical potential for $\chi$. 
The relic abundance of $\chi$ can then be found using standard techniques. The 
$\chi$ distribution departs from its new equilibrium at $x_\chi$, defined through 
\begin{equation}
	\frac{\lambda_\chi Y_\zeta(x_\chi>x_\zeta)}{x_\chi^2}\simeq 1 \, .
\end{equation}
Substituting $Y_\zeta$ in the equation above with its definition (Eq.~\eqref{eq:new_Y}), one can solve the above equation for $x_\chi$:
\begin{equation}
	x_\chi=x_{\zeta} \frac{x_{\zeta}+\sqrt{x_{\zeta}^{2}+4 \left(\frac{\langle\sigma_{\chi \zeta \to \zeta\zeta} v\rangle}{\langle\sigma_{\zeta \zeta \to \rm sm \, sm} v\rangle}\right)^{2}\left(1+x_{\zeta}\right)}}{2\left(1+x_{\zeta}\right)} \, .
\end{equation}
One can easily verify that indeed $x_\chi>x_\zeta$ for ${\langle\sigma_{\chi \zeta \to \zeta\zeta} v\rangle}>{\langle\sigma_{\zeta \zeta \to \rm sm \, sm} v\rangle} $. 
After decoupling $\chi$ the inverse process $\chi \zeta \to \zeta \zeta $ stops, and density of $\chi$ evolves as
\begin{equation}
 	\frac{d Y_{\chi}}{d x}\simeq-\frac{\lambda_\chi}{x^{2}}Y_{\zeta}(x>x_\zeta)Y_\chi(x) \, .
 \end{equation} 
 For simplicity we assume now $x_\chi\gg x_\zeta$, in that case the equation above reduces to
 \begin{equation}
 	\frac{d \log Y_{\chi}}{d x}\simeq-\frac{x_\chi^2}{x^{2}} \, .
 \end{equation}
 Integrating this equation from $x_\chi$ to $\infty$ one finds
 \begin{equation}
 	Y_\chi^\infty \simeq \tilde{Y}_\chi(x_\chi)e^{-x_\chi}\simeq Y_\zeta^\infty \exp\left[-x_\chi\left(2-r^{-1}\right)\right] \, .
 \end{equation}
 Since $Y_\zeta^\infty \sim e^{-x/r}$ but also $Y_\zeta^\infty\sim \lambda_\zeta^{-1}$, by requiring that $\chi$-radiation equality happens at $T_{\rm eq}\sim 0.8~\rm eV$ we find
 \begin{equation}
 	m_\chi\sim\left[\left(\alpha_\zeta^2m_{\rm pl}\right)^{r+\Delta}T_{\rm eq}\right]^\frac{1}{1+r+\Delta} \, ,
 \end{equation}
 with
 \begin{equation}
 	\Delta=r\frac{x_\chi}{x_\zeta}+(r-1)\left(\frac{x_\chi}{x_\zeta}-1\right)>0 \, ,
 \end{equation}
which are Eqs.~\eqref{eq:phase2} and~\eqref{eq:Delta} in the main text.

\bibliography{exception.bib}

\end{document}